\documentclass{article}

\usepackage{url}
\usepackage{ucs}
\usepackage[utf8]{inputenc}
\usepackage{amsmath}
\usepackage{mathrsfs}
\usepackage{amssymb}
\usepackage{mathtools}
\usepackage[english]{babel}
\usepackage[T1]{fontenc}
\usepackage[pdftex]{graphicx}

\usepackage[pdftex]{hyperref}

\usepackage{authblk}

\title{Combining parameter values or $p$-values}
\author[1,2]{Louis Lyons\thanks{\href{mailto:Louis.Lyons@physics.ox.ac.uk}{Louis.Lyons@physics.ox.ac.uk}}}
\affil[1]{\it \small Department of Physics, Imperial College, London SW7 2AZ, UK}
\affil[2]{\it \small Particle Physics, University of Oxford, OX1 3RH, UK}
\author[3]{\'{E}milien Chapon\thanks{\href{mailto:emilien.chapon@cern.ch}{emilien.chapon@cern.ch}}}
\affil[3]{\it \small Experimental Physics Department, CERN, CH-1211 Geneva 23, Switzerland}

\begin{document}

\maketitle

\tableofcontents

\section{Introduction}
As far as combination is concerned, it is better to go back to the original data, rather 
than to combine results.
However, this is not always possible. Also combining data can involve a very large 
amount of work.

We consider two different sorts of combination of results: parameter values and $p$-values.

\section{Parameter Values}
This can be the combination of several determinations of a single parameter, or of 
two or more parameters.
The uncertainties on the different determinations can be independent or correlated; 
and for more than one parameter, also
the parameter uncertainties of an individual determination may be independent or 
correlated.

\subsection{One parameter, no correlations}
\label{OneP_NoC}
We assume that there are N determinations $x_i\pm \sigma_i$ of a physical quantity $x$.
A way of estimating $x_{comb}$ is to minimise the weighted sum of squares
\begin{equation}
S(x) = \Sigma[(x-x_i)/\sigma_i]^2,
\end{equation}  
with the summation running over the number of observations\footnote{Many people refer to this as $\chi^2$.
We prefer to use a different symbol ($S$), as this makes more understandable the question of whether or not the distribution of $S$
is the mathematical $\chi^2$.}. 
This yields
\begin{equation}
x_{comb} = \Sigma w_i x_i/\Sigma{w_i} , \ \ \ w_i = 1/\sigma_i^2
\end{equation}
i.e. the best value of $x$ is given by the weighted average of the $x_i$, where the weights are 
equal to $1/\sigma_i^2$. Thus the smaller the uncertainty on a measurement,
the larger the weight. In an informal sense, the weight of an experiment can be 
thought of as its information content.

The uncertainty $\sigma_{comb}$ on $x_{comb}$ is given by
\begin{equation}
\label{uncertainty}
1/\sigma_{comb}^2 = \Sigma(1/\sigma_i^2)
\end{equation}
This ensures that  $\sigma_{comb}$ is at least as small as the smallest of the 
individual uncertainties; this is the motivation for combination.
It is also guaranteed to be not larger than the uncertainty on the unweighted average.
In terms of the weights, equation~\ref{uncertainty} is $w_{comb} = \Sigma{w_i}$, i.e. the
information content of the combined value is the sum of those for each of the individual 
measurements.

The uncertainties in this note depend only on the uncertainties of the individual measurements, and
their possible correlations, but {\bf not} on the degree of consistency of the separate measurements.
Thus the uncertainty on the combination of  uncorrelated measurements $0\pm 3$ and $z\pm 3$ will
be $\sim 2$, independent of whether $z$ is $1$ or $7$.
\subsection{An apparent counter-example} 
An example demonstrates that care is needed in applying the formulae. Consider high
energy cosmic rays being recorded  by a large counter system for two consecutive one-week periods, with the number of counts being
$100 \pm 10$ and $1 \pm 1$ \footnote{It  is a crime (punishable by a forcible transfer
to doing a doctorate on Astrology) to combine such discrepant measurements. It seems likely that someone turned off
the detector between the two runs; or there was a large background in the first measurement which was eliminated
for the second; etc. The only reason for using such discrepant numbers is to produce a dramatically stupid
result. The effect would have been present with measurements like $100 \pm 10$ and $81 \pm 9$.}.
Unthinking application of the formulae for the combined result gives the ridiculous $2 \pm 1$. What has gone wrong?

The answer is that we are supposed to use the {\bf true} accuracies of the individual measurements to assign the weights. 
Here we have used the {\bf estimated} accuracies. Because the estimated uncertainty 
depends on the estimated rate\footnote{The problem arises here because the standard deviation in a Poisson process is equal
to the square root of the rate. Even worse, in determining the lifetime $\tau$  of a particle from a set of $N$ measured decay times, the uncertainty on $\tau$ is $\tau/\sqrt N$ i.e. the estimated uncertainty is proportional to the estimated lifetime.}, 
a downward fluctuation in the measurement results in an underestimated uncertainty,
an overestimated weight, and a downward bias in the combination. In our example, the combination should assume 
that the true rate was the same in the two measurements which used the same detector and which 
lasted the same time as each other, and hence their
true accuracies are (unknown but) equal. So the two measurements should each be given the same weight, which 
yields the more sensible combined result of $50.5  \pm 5$ counts per week. 

A general way of mitigating this problem within this approach is discussed in ref~\cite{LMS}. It incorporates the way the uncertainty
for each result is  expected to vary with the estimated parameter value. 
It is equivalent to an iterative approach, in which at each stage, the input uncertainties are recalculated assuming that they 
can be obtained using the parameter value as determined in
the previous iteration.        

\subsection{{\bf BLUE} for one parameter with correlated measurements}
A method of combining correlated results is the `{\bf B}est {\bf L}inear {\bf U}nbiassed {\bf E}stimate' ({\bf BLUE})~\cite{LGC}. 
We look for the best linear unbiassed combination
\begin{equation}
x_{BLUE} = \Sigma w_i x_i,
\end{equation} 
where the weights are chosen to give the smallest uncertainty $\sigma_{BLUE}$ on 
$x_{BLUE}$. Also for the combination to be unbiassed, the weights must add up to unity.
They are thus determined by minimising $\Sigma\Sigma w_i w_j E^{-1}_{ij}$, subject to the constraint
$\Sigma w_i = 1$; here $E$ is the covariance matrix for the correlated measurements. This gives
\begin{equation}
w_i = \Sigma e_{ij}/\Sigma\Sigma e_{ij},
\end{equation}
where $e_{ij}$ is an element of $E^{-1}$, and the summation in the numerator is over the index {j}, while
the double summation in the denominator is over $i$ and $j$.

The $BLUE$ procedure just described is equivalent to the $\chi^2$ approach for checking whether
a correlated set of measurements are consistent with a common value. The advantage of $BLUE$ is that 
it provides the weights for each measurement in the combination. It thus enables us to calculate the contribution 
of various sources of uncertainty in the individual measurements to the uncertainty on the combined result.

When the correlation is so strong that the correlation coefficient\footnote{Here 
$\sigma_1 \leq \sigma_2$, and  $\rho$ is the 
covariance divided by the product $\sigma_1\sigma_2$.} 
$\rho > \sigma_1/\sigma_2$, the best estimate of $x$ falls outside the range of $x_1$
and $x_2$. This is in fact reasonable. If the correlation is strongly positive, it is likely that $x_1$ 
and $x_2$ lie on the same side of the true value $x_{true}$, with $x_1$ (the measurement with the smaller uncertainty)
lying closer to $x_{true}$ than $x_2$ does. Thus it is entirely sensible that the best estimate should
involve extrapolating from $x_2$ to beyond $x_1$. 

However, the resulting $x_{BLUE}$ is sensitive to the 
values of the uncertainties and the correlations, so combining  highly correlated values may 
not be sensible. This situation can arise, for example, when there is more than one group
within a collaboration, analysing more or less the same data but with slightly different 
analyses, and with the same physics aim, e.g. measuring the top quark mass. This is likely to produce a set of answers that 
will be highly correlated. Rather than trying to combine the different results, it is better to decide which 
procedure should be used as the published result of the Collaboration (with the others being simply confirmatory).
This choice should be based {\bf not} on the results of the different methods, but rather on the expected sensitivity of 
each method; etc. 

Another feature of large correlations is that the uncertainty on the combined value tends to zero as the correlation 
coefficient tends to +1 or -1. (Remember that even with complete correlation, the uncertainties do not have to be equal.) 

When the individual measurements are uncorrelated, ${\bf BLUE}$ simplifies to the method described in Section~\ref{OneP_NoC}.

\subsection{Why weighted averaging can be better than simple averaging}
Consider a remote island whose inhabitants are very conservative, and no-one leaves or arrives 
except for some anthropologists who wish to determine the number of married people there.
Because the islanders are very traditional, it is necessary to send two teams of anthropologists,
one consisting of males to interview the men, and the other of females for the women. There are too
many islanders to interview them all, so each team interviews a sample and then extrapolates. 
The first team estimates the number of married men as $10,000 \pm 300$. The second, who 
unfortunately have less funding and so can interview only a smaller sample, have a 
larger statistical uncertainty; they estimate $9,000 \pm 900$ married women. Then how many 
married people are there on the island? 

The simple approach is to {\bf add} the numbers of married men and women, to give $19,000 \pm 950$
married people. But if we use some theoretical input, maybe we can improve the accuracy of 
our estimate. So if we assume that the islanders are monogamous, the true numbers of married men and 
women should be equal. The {\bf weighted} 
average is $9,900 \pm 285$ married couples and hence $19,800 \pm 570$ married people.

The contrast in these results is not so much the difference in the estimates, but that 
incorporating the assumption of monogamy and hence using the weighted average gives a smaller 
uncertainty on the answer. Of course, if our assumption is incorrect, this answer will be biassed.

A Particle Physics example incorporating the same idea of theoretical input reducing the 
uncertainty of a measurement is `Kinematic Fitting'. There the uncertainties on the measured momenta 
and energies of the objects produced in high energy interactions are reduced by assuming that 
energy and momentum conservation applies between the initial state collision particles and the final state objects 
measured in the detector.

\section{Two or more  Parameters}
\subsection{Different measurements uncorrelated}
 
There are situations where analyses determine two or more parameters. For example:
\begin{itemize}
\item{When fitting a peak plus a smooth background to a mass spectrum, the parameters 
include the location  and strength of the signal, and perhaps its width. }
\item{In a search for neutrino oscillations where only two flavours are relevant, the parameters are the amplitude of the oscillations 
$\sin^2 2\theta$; and $\Delta m^2$, the difference of the mass-squared of the two neutrinos, which determines the oscillation frequency.}
\item{One or more physics parameters $\phi$ and nuisance parameter(s) $\nu$ for systematic effects.}
\item{Straight line fitting. The parameters are the gradient and the intercept of the line.}
\end{itemize}
In many cases the uncertainties  on the parameters will be correlated. 

\vspace{0.1in}
When several independent\footnote{If there are correlations among the $N$ separate measurements of the two parameters, 
 the covariance matrix $E$ is expanded to be of size $2N \times 2N$, and equation 
(\ref{uncorrelated_correlated}) is readily modified to include all correlations -- see Section~\ref{Valassi}. }    
measurements of the correlated parameters exist, we may want to combine the results. 
For a pair of parameters, the weighted sum of squares for consistency with a particular $(a,b)$ is
\begin{equation}
\label{uncorrelated_correlated}
S(a,b)= \Sigma [(a_i-a)^2 e_{i,11} + (b_i-b)^2 e_{i,22}  + 2(a_i-a)(b_i-b) e_{i,12}]
\end{equation}
where the summation is over the $N$ independent measurements $(a_i, b_i)$,
and $e_{i,kl}$  is the $(k,l)$ element of the inverse covariance matrix $E^{-1}_i$ of the $i^{th}$ measurement. 
Then $S(a,b)$ is simply minimised with respect to the parameters $a$ and $b$. The uncertainties and correlation
for the combined values are given by the covariance matrix $M$; the elements of  its inverse are 
\begin{equation}\begin{split}
\label{matrix}
M^{-1}_{11}& = 0.5*\frac{\partial^2 S}{\partial a^2}  \\
M^{-1}_{22}& = 0.5*\frac{\partial^2 S}{\partial b^2}  \\
M^{-1}_{12}& = 0.5*\frac{\partial^2 S}{\partial a \partial b}
\end{split}
\end{equation}

The extension to the case where each analysis measures more than two parameters is straightforward.

 \subsection{Different  measurements correlated}
 \label{Valassi}
An extension of the above example is where we have $N$ observables, each of which is measured in $p$ 
different experiments, and there are possible correlations in all $n = N \times p$ variables. Valassi has extended BLUE,
using the criterion of minimising the uncertainty on each of the $N$ combined values~\cite{Valassi:2003mu}.
For Gaussian uncertainties, this is shown to be equivalent to minimising the weighted sum of squares. 
An example where this might be used 
would be a measurement of the differential cross section for some process, using several different decay channels;
there could be correlations across the bins of the cross-section for a given channel, and also among the different channels.
The output is the differential cross-section, with each bin being the combination of the different channels, taking all
correlations into account. 

The Valassi procedure is as follows.
 Let us assume that we are interested in $N$ observables, $X_\alpha = \{X_1, ..., X_N\}$, and that we have $n$ experimental results 
 $y_i = \{y_1, ..., y_n\}$, such that each of the measurements $y_i$ corresponds to one of the observables $x_\alpha$ (and all observables are
 measured at least once: $n \geq N$). The $(n \times N)$ matrix $\mathscr{U}$ is defined by
 \begin{equation}
  \mathscr{U}_{i\alpha} = \left\{ \begin{matrix*}[l]
  1 & \textrm{if } y_i \textrm{ is a measurement of } X_\alpha, \\
  0 & \textrm{if } y_i \textrm{ is not a measurement of } X_\alpha.
 \end{matrix*} \right.
 \end{equation}\\
Each of the $n$ rows of $\mathscr{U}$ has one and only one element equal to 1. For instance, if we would combine a 3-bin differential
 cross section measurement between two channels, e.g. muon and electron, then $n = 6$, $N = 3$, and the $\mathscr{U}$ matrix would be:
 
 \begin{equation}
  \mathscr{U}_{i\alpha} = 
  \begin{pmatrix}
  1 & 0 & 0 \\
  0 & 1 & 0 \\
  0 & 0 & 1 \\
  1 & 0 & 0 \\
  0 & 1 & 0 \\
  0 & 0 & 1
  \end{pmatrix}
 \end{equation}

 Let us also define the $(n \times n)$ covariance matrix of
 the measurements,
 
 \begin{equation}
  \mathscr{M}_{ij} = \textrm{cov}(y_i, y_j).
 \end{equation}

 Then the best linear estimate of each observable $X_\alpha$ is
 
 \begin{equation}
  \hat{x}_\alpha = \sum_{i=1}^n \lambda_{\alpha i} y_i,
 \end{equation}\\
 where the weights $\lambda_{\alpha i}$ are
 
 \begin{equation}
  \lambda_{\alpha i} = \sum_{\beta=1}^N \left( \mathscr{U}^t \mathscr{M}^{-1} \mathscr{U} \right)^{-1}_{\alpha\beta} 
 \left(\mathscr{U}^t \mathscr{M}^{-1} \right)_{\beta i},
 \end{equation}\\
 and $\mathscr{U}^t$ is the transpose matrix of $\mathscr{U}$. The covariance matrix for the estimates is
 
 \begin{equation}
  \textrm{cov}(\hat{x}_\alpha, \hat{x}_\beta) = \left( \mathscr{U}^t \mathscr{M}^{-1} \mathscr{U} \right)^{-1}_{\alpha\beta}.
 \end{equation}

\subsection{Detailed Example: Straight Line Fits}

\begin{figure}[ht]
\begin{center}
\includegraphics[width=\textwidth]{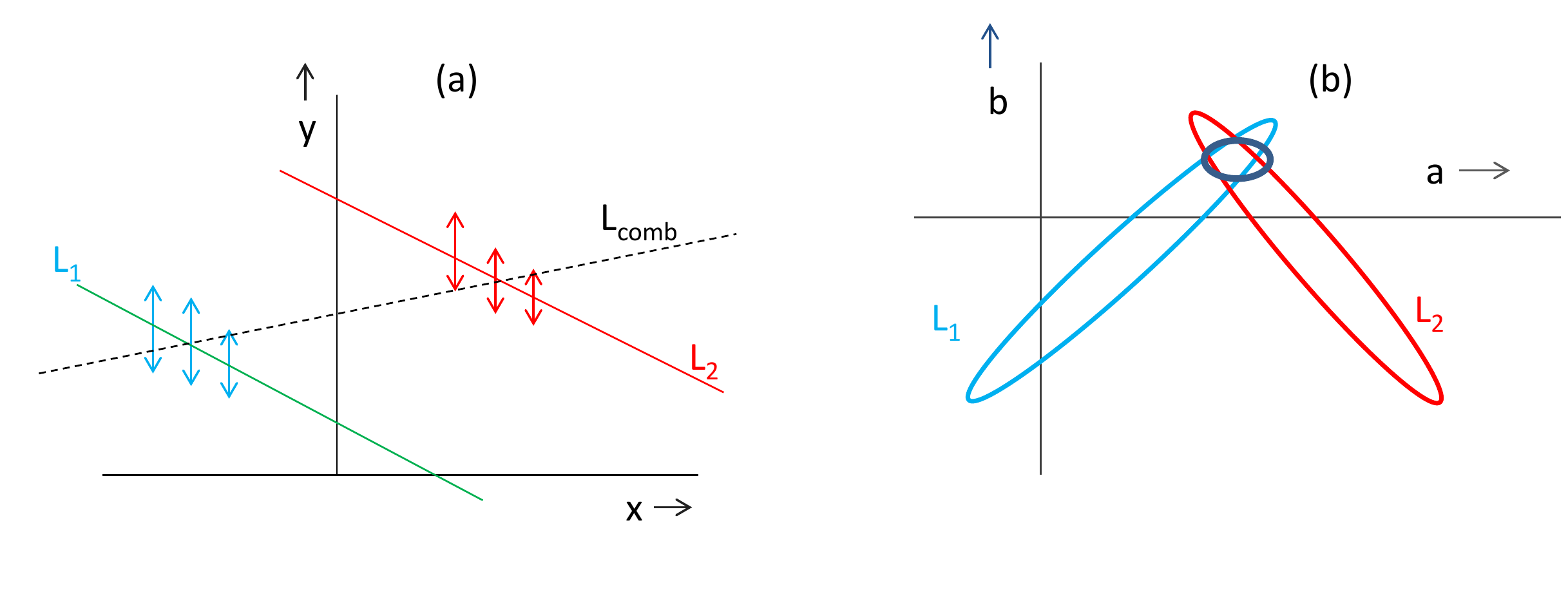}
\caption{(a) Three hits in each of 2 sub-detectors. The line $L_1$ is a fit to the three left-most points,
and $L_2$ is for the three on the right. $L_{comb}$ is the result of correctly combining the intercepts
and gradients ($a$ and $b$ respectively) for $L_1$ and $L_2$, taking the correlations between $a$ and $b$ 
into account; or equivalently, of fitting a line to all 6 hits. (b) The large covariance ellipses for $L_1$ and $L_2$,
and the small one for $L_{comb}$. The big improvement from combining is dramatically evident.
\label{tracking1}}
\end{center}
\end{figure}

Here we discuss a very simple example of combining straight line fits, where the answer can be appreciated intuitively.
It consists of a simplified tracking situation, in which a particle passes through 6 detector planes, of which 3 are closely spaced 
and separated by some distance from another 3 closely spaced planes (See Fig.~\ref{tracking1}). 
The data consist of independent measurements $y_i \pm \sigma_i$ at well-defined $x_i$ values. There is no magnetic field and we consider only 
the $x$ and $y$ coordinates so the track is parametrised by the straight line $y=a+bx$.  A straight line $L_1$ is fitted to the hits in the 
3 left-most planes, and $L_2$ to the 3 right-most planes. Finally the results $(a_1,b_1)$ and $(a_2,b_2)$ are combined to give
$(a_{comb},b_{comb})$.

A straight line fit to a set of closely separated points will determine the line's gradient $b$ with a large uncertainty. Furthermore, 
if these points are centred away from $x=0$, there will be a strong correlation between $a$, the intercept at $x=0$, and $b$.
The large uncertainty on the gradient $b$ then results in a large
uncertainty on the intercept $a$. 
The covariance of $a$ and $b$ is obtained from equations~\ref{matrix}, and is proportional to $-\langle x\rangle$, 
where $\langle x \rangle$ is the weighted average of the $x$-positions of the fitted data points (i.e. $\Sigma(x_i/\sigma_i^2)/\Sigma(1/\sigma_i^2)$). 

Thus we expect that the lines $L_1$ and $L_2$  will have large uncertainties on $a$ and $b$, and strong correlations but 
of opposite signs. However,  because of the larger range of $x$-values, $L_{comb}$ will have very much smaller uncertainties.
The covariance ellipses for $ L_1$ and $L_2$, as well as for $L_{comb}$ are shown in Fig.~\ref{tracking1}, and bear out these expectations. 
In terms of the covariance  ellipses shown there, it is because they have different orientations that the combination results in 
vastly reduced uncertainties. 

\begin{figure}[h!t]
\begin{center}
\includegraphics[width=\textwidth]{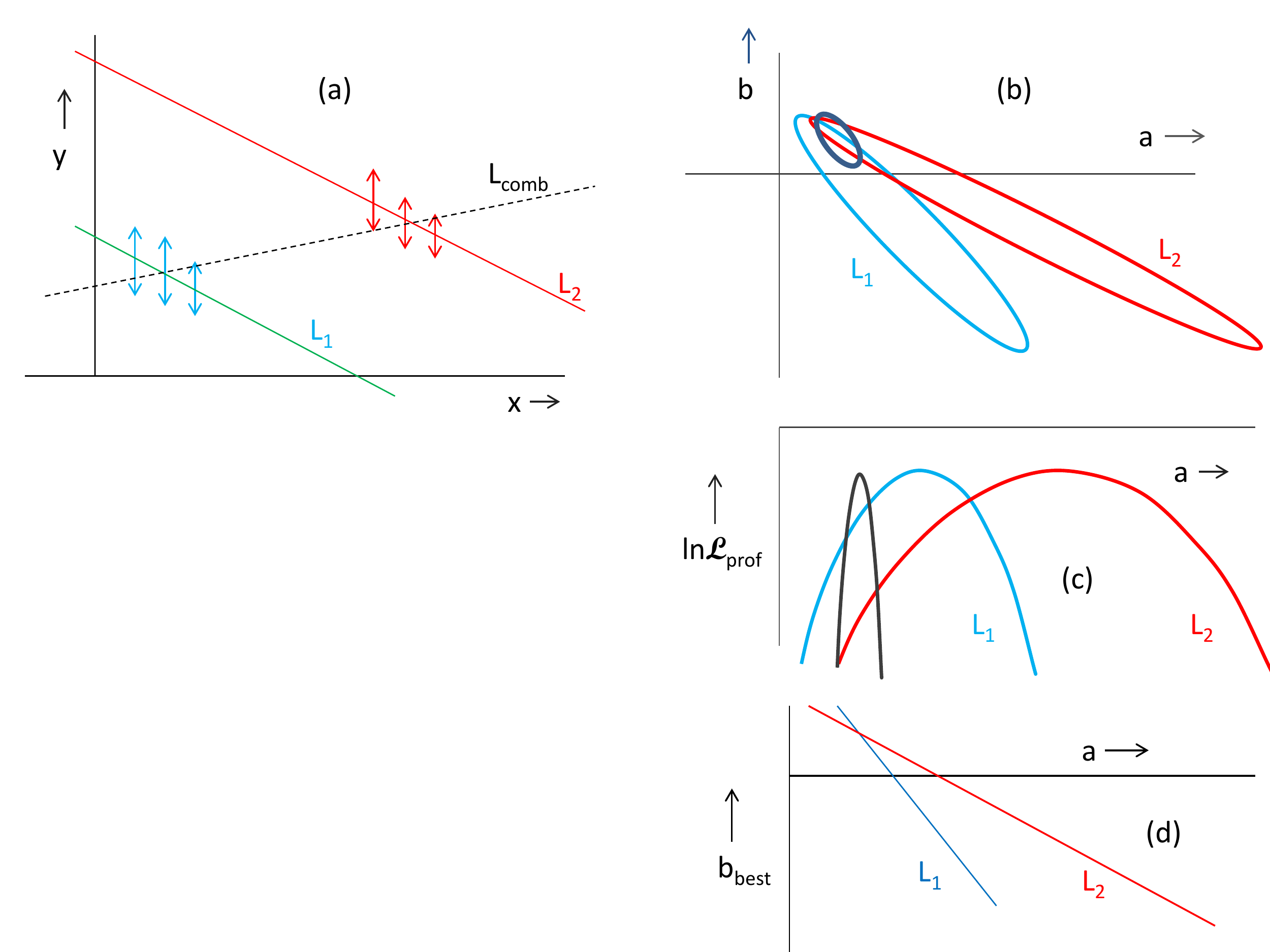}
\caption{(a) and (b) As in Fig.~\ref{tracking1} but the two sets of hits are now on the same side of the vertical axis. 
In this example, the best
values of  both the intercept and the gradient of the combination lie outside the individual values 
for $L_1$ and $L_2$. (c) The logarithm of the profile likelihoods for $L_1$, $L_2$ and $L_{comb}$, as functions of $a$. 
The incorrect procedure of combining the profile likelihoods for $L_1$ and $L_2$ does not give that for  $L_{comb}$;
it would result in a parabola slightly narrower than that for $L_1$, and with its maximum a little to the right 
of $L_1$'s maximum.
(d) The best value of $b$ as a function of $a$ for the two lines $L_1$ and $L_2$. The overall best values of $b$ 
for the two lines are equal; however,  except when $a = a_{comb}$, the values of $b_{best}(a)$ for these 
lines do not agree. That
explains why combining  the profile likelihoods for the two lines is not a sensible procedure.
\label{tracking2}}
\end{center}
\end{figure}

When the two sets of sub-detector planes are centred on the same side of the origin, as is usually the case in tracking,
the best values of both the gradient and of the intercept of the combined line can be outside the ranges of the corresponding quantities for lines $L_1$ and $L_2$ (see Fig.~\ref{tracking2}).

For the straight line fits, the results for $(a,b)$ and for their covariance matrix $M$ are the same whether we combine the $L_1$ and $L_2$ 
results, or whether we do a single fit to all 6 data points.

\begin{figure}[h!t]
\begin{center}
\includegraphics[width=\textwidth]{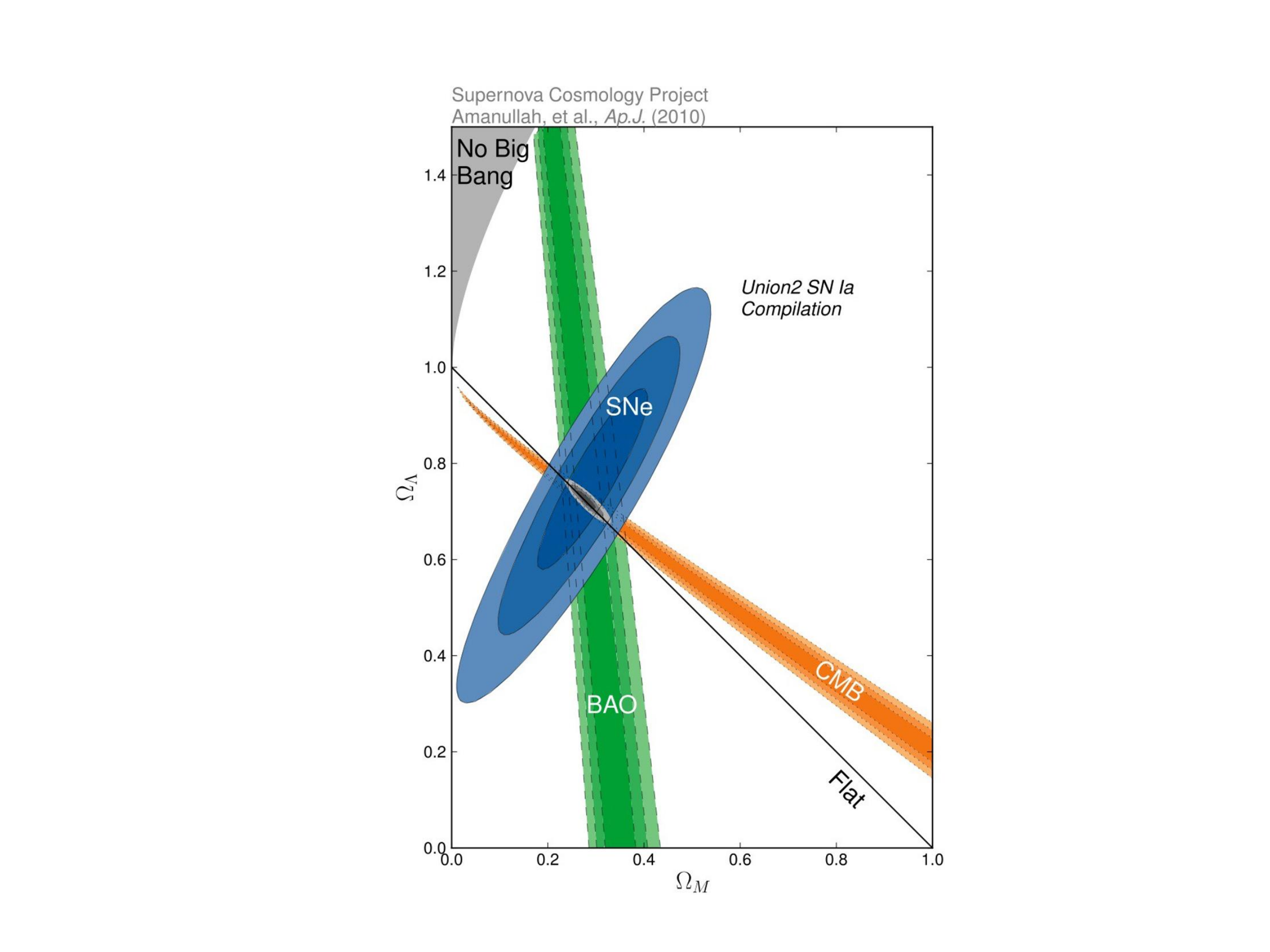}
\caption{Confidence regions for $(\Omega_{\Lambda}, \Omega_M)$, the fractions of Dark Energy and 
Dark Matter respectively, as determined from Supernovae (SNe), the Cosmic Microwave Background (CMB) 
and Baryon Acoustic Oscillations (BAO). Each individual measurement has a large uncertainty on 
$\Omega_{\Lambda}$, but because the methods have  different correlations, the combination 
determines it well.
\label{dark}}

\end{center}
\end{figure}

 A physical example of the big reduction in uncertainty when combining results of pairs of parameters with
large internal correlations  is the determination of the fraction of dark energy in the Universe $\Omega_{\Lambda}$.
There are several different methods that provide information on this and the fraction of dark matter $\Omega_M$, 
but each on its own has a large uncertainty. 
Because of their different correlations, however, their combination  provides a precise measurement of $\Omega_{\Lambda}$
(see Fig.~\ref{dark}).

\subsection{Profile Likelihood}
For situations where we have several parameters, it is common and sometimes natural to choose 
one as the parameter of interest ($\phi$) and to
profile or to marginalise over all the others ($\nu$); this is especially common when we have one physics parameter, and several 
nuisance parameters related to systematic effects. The profile likelihood is 
\begin{equation}
 \mathcal{L}_{prof}(\phi) = {\mathcal L}(\phi,\nu_{best}(\phi))     
 \end{equation}
where $\nu_{best}(\phi)$ is the value for $\nu$ that maximise the likelihood at that particular $\phi$.
The profile likelihood is thus a function just of $\phi$. Fig.~\ref{tracking2}(c) shows the profile likelihood for the intercept $a$ (i.e. profiled
over the gradient $b$) for all three lines. The widths of the curves for $\ln\mathcal{ L}_{prof}$ correctly give the uncertainties on $a$.
However it is important to note that, in contrast to the situation with the full likelihoods $\mathcal{ L}(a,b)$,           
combining the profile likelihoods for $L_1$ and $L_2$ would
 {\bf not} give the profile likelihood for $\mathcal{L}_{comb}$,
even though the best values of  $b$ for the two lines happen to be the same.  
However, except when $a = a_{best}$, the values of 
$b_{best}(a)$ for $L_1$ and for $L_2$ at the same $a$ are different  - see Fig.~\ref{tracking2}(d)); 
that is a reason that the combination of profile likelihoods is not a sensible procedure.

\subsection{Better combination?}
If the only information available is the set of values $\mu_i$ from the separate  measurements and their covariance matrix ,
then the only possibilities for combining are the methods described above. With a little 
more information, however, it might be possible to reconstruct approximately the likelihood
functions and then to combine them rather than just the results.

For example, when the uncertainties are asymmetric, Barlow has suggested various ways  of modifying the Gaussian shape
by prescriptions for how the width varies~\cite{Barlow}. 

A note by Cousins~\cite{log_normal} points out that in some simplified circumstances, the shape of the likelihood is defined. Thus if 
the result is obtained by multiplying several factors, the equivalent of the central limit theorem causes the distribution of the product to be approximately log-normal. 
Some systematic uncertainties might also result in such a distribution. For example, theorists might say
that their predicted cross-section for some process was accurate to within a factor of 2.

Alternatively for a search for a signal
involving Poisson counting in the signal and in background regions (the `on-off' problem), the distribution is a gamma function. This also 
applies to lifetime determinations using individually observed decay times with an expected exponential distribution. In all cases the parameters 
of the expected distributions are determined from the numerical values of $\mu$ and $\sigma$. 

Of course the ideal situations 
for these distributions to be relevant are rarely realised in practice. For example, for the gamma distribution to apply to  the lifetime measurement,
we require the expected decay distribution to follow a perfect exponential. This means that we have a constant efficiency for observing decays
over the full range of decay times from zero to infinity, and can ignore backgrounds, time resolution, etc.

\subsection{Varying $\rho$}
Because correlations can  lead to extrapolation and to small uncertainties, it is sometimes suggested that it
would be a good idea  to set the correlation coefficient $\rho$ to zero.
This is thought to be conservative, but it  throws away information and is against the spirit of {\bf BLUE} - 
 `B' stands for `Best', which means `{\bf smallest} uncertainty', rather than `most conservative'.
Furthermore $\rho = 0$ is not the  most conservative choice. For example, if we have analysed a large 
data set $D$ and also a subset $S$ of this data, and we combine them using the covariance matrix with elements
\begin{equation}
E_{11} = \sigma_D^2, \ \ \ E_{22} = \sigma_S^2, \ \ \ E_{12} = E_{21} = \sigma_S^2,
\end{equation}
the weight ascribed to $S$ turns out to be zero, i.e. the subsample $S$ is ignored 
and the `combined' result is simply that from $D$ (as is sensible). 
But if we set the covariance to zero, our combined result will have an incorrect `improvement'
with reduced uncertainty $\sigma_C$ given by
\begin{equation}
1/\sigma_C^2 = 1/\sigma_D^2 + 1/\sigma_S^2
\end{equation}

While it might be sensible to choose the most conservative value 
for $\rho$ when the value of $\rho$  is unknown, otherwise its actual value seems a better choice.

Another problem is that in ignoring $\rho$, we will obtain an incorrect contribution to the weighted sum of squares $S$. 
Thus if the $pdf$ in $(x,y)$ is a 2-dimensional Gaussian centred on the origin with $\sigma_x = \sigma_y =1$ and $\rho = +0.9$, 
the correct contribution to $S$  from a measurement at (+1.0, +1.0) is ~1, while for (+1.0, -1.0) it is 20; setting $\rho=0$ would 
incorrectly result in a contribution of 2 for both of them.

\section{$p$-values}

Sometimes the effect of  New Physics could appear in several different reactions in a given experiment; or in 
different experiments. 
It would then be sensible to combine the information, in order to improve the sensitivity of the search. The best way 
to do this is to perform a joint analysis, but this is not always possible, so an alternative is to combine the individual 
$p$-values. Problems with this are:
\begin{itemize}
\item{Different effects: Two analyses might each have small $p_0$-values because they both disagree with the Standard Model,
but have different inconsistent  discrepancies e.g. peaks at quite different mass values.}
\item{Non-uniqueness: Bob Cousins~\cite{Cousins} has pointed out that in the combination of $N\ p$-values, we are trying to find a 
transformation from the $N$ (presumed uniform) 1-dimensional distributions\footnote{The assumption of a uniform distribution for $p$-values will not be true if the data are discrete.} to just one uniform 
distribution; this 
can clearly be achieved  in many ways, thus yielding a large variety of different possible combined $p$-values. 
Which is best requires extra information about the possible alternative hypotheses, and also more details of the analyses 
beyond just their individual $p$-values. Again the desirability of a combined analysis is demonstrated.}  
\item{Selection bias: Care must be taken to combine all relevant analyses, and not just the ones which give small 
individual $p$-values.}
\end{itemize}

One possibility is to calculate the product $z_{obs}$ of the $n$ individual $p$-values. Then the probability $P$ of the product
of $n$ independent uniformly distributed $p$-values being smaller than $z_{obs}$ is
\begin{equation}
\label{combine_by_product}
P = z_{obs} \Sigma (-\ln z_{obs})^j/j!
\end{equation}
where the summation extends over $j = 0$ to $n-1$. Thus $P$ is larger than $z_{obs}$.
For two measurements $P = p_1p_2(1-\ln (p_1p_2))$.

If the $p_i$-values were obtained from weighted sums of squares $S_i$ 
which are expected to have $\chi^2$ distributions with numbers of degrees of 
freedom $\nu_i$, an alternative is to use $S_{comb} = \Sigma S_i$ and 
$\nu_{comb} =\Sigma \nu_i $ to obtain the overall $P$. In the special case where
the individual $\nu_i$ are all 2, this becomes equivalent to using eqn. \ref{combine_by_product}. 

A third approach is the Stouffer method~\cite{Stouffer} which uses
\begin{equation}
z_{comb} = \Sigma z_i/\sqrt n,
\end{equation}
where $z_i$ are the signed $z$ scores (i.e the number of standard deviations corresponding to the one-sided $p$-value, 
with $p = 0.5$ being equivalent to $z = 0$) and $z_{comb}$ is the combined value.

\section{Conclusion} 

\begin{table}
\caption{Summary Table. For all cases below, beware of using {\bf estimated} uncertainties and correlations.}
\begin{center}
\begin{tabular} {  | c  |  c | c | c |}
\hline
No.  of params    &   Correlated?   &      Method    &   Result  \\   \hline
1      &      No      &    $\chi^2$  &  $w_i = 1/\sigma_i^2$    \\      \hline
1       &      $\rho \neq 0$    &   $\chi^2$  & If $\rho > \sigma_1/\sigma_2, x_{comb}$ uses extrapolation      \\
        &                                 &   BLUE       & Gives weights for each $x_i$  \\    \hline
2 $(x,y)$       &   $x$ and $y$ correlated     &    $ \chi^2$           & $(x,y)_{comb}$ can be out of range of ($x,y)_i$ \\  
                              &                                               &                               &  $(\sigma_x, \sigma_y)_{comb}$ can be much smaller than 
$(\sigma_x,\sigma_y)_i$   \\ \hline
\hline
 n $p$-values     &   No   &     Many     &   Best method requires more than just $p$-values  \\ \hline
\end{tabular}
\end{center}
\label{Summary_Table}
\end{table}

Although it is useful to combine results of different measurements of the same physical parameter(s), 
it is almost always better to perform a single combined analysis of all the data. 
However, for cases where this is not possible or is impractical, we discuss here combinations for the 
measurement of a single parameter; of two or more parameters; and of $p$-values. For parameter determination, the simpler situation  is where the
individual uncertainties are uncorrelated; we also discuss  the correlated case.

Large correlations can result in the combined value lying outside the range of the individual values; and to significantly reduced uncertainties.  This is not unreasonable, but can be dangerous if the uncertainties and/or correlations are inaccurately estimated. However, setting correlations to zero can
result in throwing away important information.

The main results of combinations are summarised in the Table. 

\section{Acknowledgments}

We wish to thank Bob Cousins and Andrea Valassi  for enlightening conversations, and to Olaf Behnke for his comments 
on an earlier version of this note.

\appendix

\section{An aside on profile likelihoods}
There is sometimes confusion on whether profile likelihood ratios involving two hypotheses
are the ratio of the profile likelihoods, or are the likelihood ratio profiled with respect 
to the nuisance parameter(s). The latter is {\bf not} a sensible procedure in that the profile 
likelihoods for the two hypotheses can require different values of the nuisance parameter(s)
at a given value of the parameter of interest.

\begin{figure}[htb]
\begin{center}
\includegraphics[width=\textwidth]{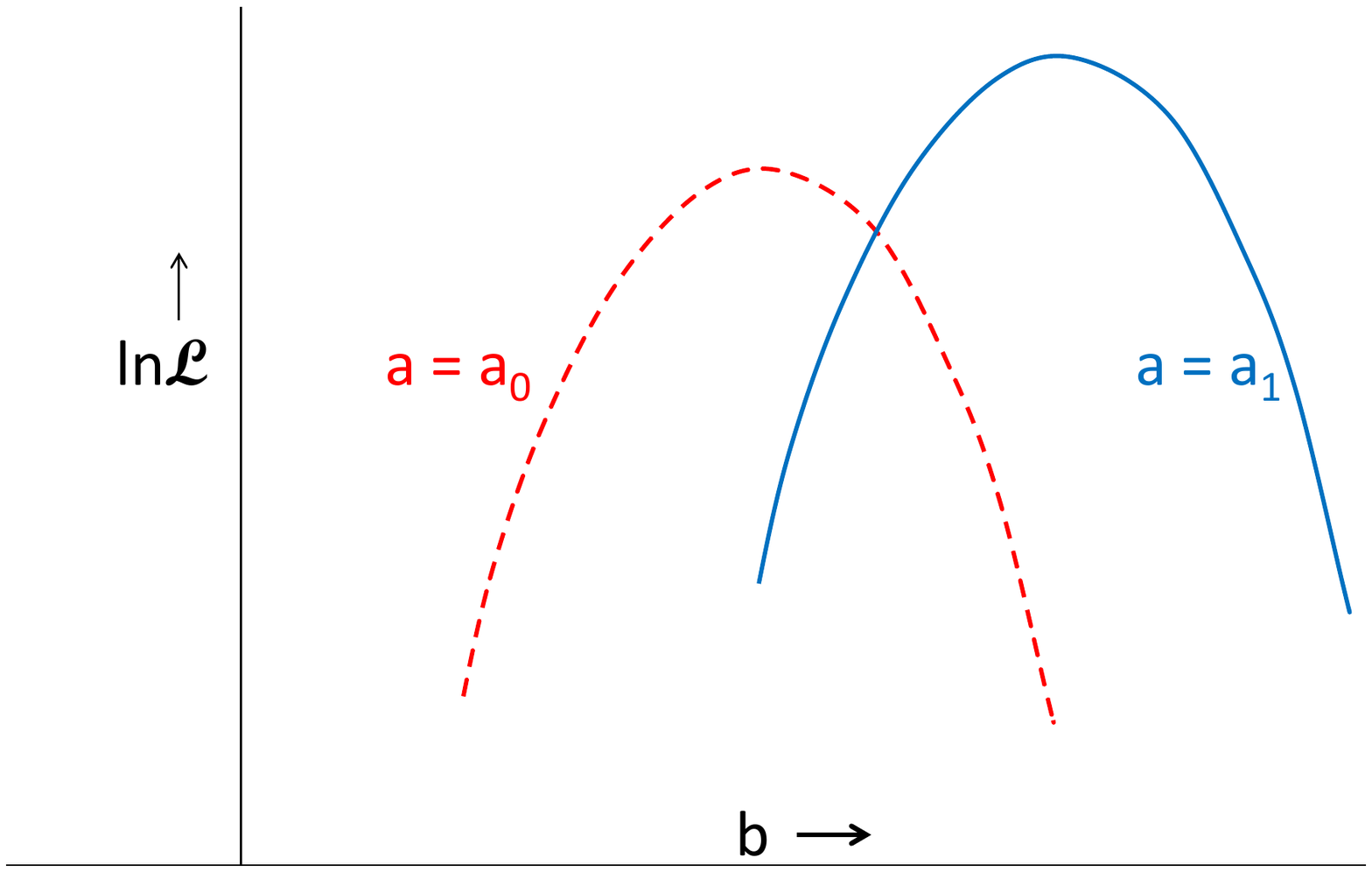}
\caption{Plots of the log-likelihoods as functions of the gradient $b$, for straight lines
with intercept $a = a_0$  or $a = a_1$. With the functions being equal width parabolae,
the log-likelihood ratio would be linear in $b$ and hence have no stationary value.
\label{parabolae}}
\end{center}
\end{figure}

For the example of a straight line fit to some data (e.g. all 6 points of Fig.~\ref{tracking1}), the 
parameter of interest might be the intercept $a$, with the gradient $b$ being
a nuisance parameter. Then the 
hypothesis test could involve two different sets of straight lines, the first with $a = a_0$
and the second with $a = a_1$. In this simple case  (and for a variety of other problems 
too), the log-likelihoods as functions of $b$ are parabolae of equal width, but with different
locations and heights of their minima (see Fig.~\ref{parabolae}). Then the difference of the 
log-likelihoods is linear in $b$, with no stationary value anywhere. Even worse, if the widths 
of the individual log-likelihoods for the two hypotheses are slightly different (as would be 
the case when the uncertainties $\sigma_i$ on the original data $y_i$-values depended on the 
parameters $a$ and/or $b$), the plot of the log-likelihood ratio would have a weak 
quadratic dependence on $b$,  so that profiling could result in a stationary value at a 
large and irrelevant value of $|b|$.

Thus profiling a ratio of likelihoods is not a good procedure.


\begin{thebibliography}{99}
\bibitem{LMS} L.  Lyons, A. J. Martin and D. H. Saxon, `On the determination of the B lifetime by combining the results of different experiments',
Phys Rev {\bf D 41} (1990) 982.
\bibitem{LGC} L.  Lyons, D Gibaut and P. Clifford, `How to combine correlated estimates of a single physical  quantity', 
Nucl Inst Meth, {\bf A270} (1988) 110.
 \bibitem{Valassi:2003mu} A.~Valassi,
  `Combining correlated measurements of several different physical quantities',
  Nucl.\ Instrum.\ Meth.\ A {\bf 500} (2003) 391.
  \url{doi:10.1016/S0168-9002(03)00329-2}
\bibitem{Cousins} R. D. Cousins,  `Annotated Bibliography of Some Papers on Combining Significances or $p$-values' (2008)
\url{https://arxiv.org/pdf/0705.2209.pdf} 
\bibitem{Stouffer}S. A. Stouffer  et al, `The American Soldier' (1949) Princeton University Press.
\bibitem{Barlow} R.  Barlow, `Asymmetrical errors', Proceedings of PHYSTAT2003, Stanford Linear Accelerator Center (2003)  \url{arXiv:physics/0401042}; and `Asymmetric statistical errors',
(2004) \url{arXiv:physics/0406120 [physics.data-an]}
\bibitem{log_normal} R. Cousins, `Probability Density Functions for Positive Nuisance Parameters' (2010) \url{http://www.physics.ucla.edu/~cousins/stats/cousins_lognormal_prior.pdf}
\end{thebibliography}
\end{document}